\documentclass[article,amsmath,11pt,amssymb,floatfix,superscriptaddress,
showpagenumber,nofootinbib]{revtex4}
\usepackage{graphicx}
\usepackage{epsfig}
\usepackage{bm}
\usepackage{amsfonts}

\input epsf

\begin{document}
\title{\Large Cosmology of a generalized version of holographic dark energy in presence of bulk viscosity and its inflationary dynamics through slow roll parameters}
\author{Gargee Chakraborty}
\affiliation{Department of Mathematics, Amity University, Major
Arterial Road, Action Area II, Rajarhat, New Town, Kolkata 700135,
India.}
\author{Surajit Chattopadhyay}
\email{schattopadhyay1@kol.amity.edu; surajitchatto@outlook.com}
\affiliation{ Department of Mathematics, Amity University, Major
Arterial Road, Action Area II, Rajarhat, New Town, Kolkata 700135,
India.}

\date{\today}

\begin{abstract}
The present study reports a reconstruction scheme of a Dark Energy (DE) model with higher order derivative of Hubble parameter, which is a particular case of Nojiri-Odintsov holographic DE \cite{NO2} that unifies phantom inflation with the acceleration of te universe on late-time. The reconstruction has been carried out in presence of bulk-viscosity, where the bulk-viscous pressure has been taken as a function of Hubble parameter. Ranges of cosmic time $t$ have been derived for quintessence, cosmological constant and phantom behaviour of the equation of state (EoS) parameter. In the viscous scenario, the reconstruction has been carried out in an interacting and non-interacting situations and in both the cases stability against small perturbations has been observed. Finally, the slow roll parameters have been studied and a scope of exit from inflation has been observed. Also, availability of quasi exponential expansion has been demonstrated for interacting viscous scenario and a study through tensor to scalar ratio has ensured consistency of the model with the observational bound by Planck. Alongwith primordial fluctuations the interacting scenario has been found to generate strong dissipative regime.
\end{abstract}

\maketitle

\section{Introduction}
Riess et al.\cite{obs1} and Perlmutter et al.\cite{obs2} independently reported in the late 90's that the current universe is passing through a phase of accelerated expansion. Their discovery is a breakthrough in the field of Modern Cosmology. They \cite{obs1,obs2} discovered this accelerated expansion by accumulating the observational data of distant Supernovae Ia (SNeIa) and their discovery has further been supported by other observational studies \cite{obs3,obs4,obs5,obs6,obs7}. Some exotic matter characterised by negative pressure is thought to be responsible for driving this acceleration. This exotic matter is dubbed as "Dark Energy" (DE) \cite{DE1,DE2}. The DE differs from the ordinary matter in the sense that it is characterised by negative pressure. A DE is described by the equation of state (EoS) parameter defined as $w=\frac{p}{\rho}$, where $p$ is the pressure and $\rho$ is the density due to DE. From Friedmann's equations one can easily verify that $w<-\frac{1}{3}$ is a necessary condition for the accelerated expansion of the universe. Cosmological constant ($\Lambda$), characterised by constant EoS parameter $w=-1$, is the simplest candidate for DE \cite{DE7}. Although $\Lambda$ is consistent with observations, other candidates of DE have also been proposed in the literature and those candidates have time varying EoS parameter. Such candidates have been proposed in the literature to get rid of some limitations of cosmological constant \cite{obs10}. Candidates with dynamic EoS parameter can be broadly differentiated into (i) scalar field models; (ii) holographic models of DE; (iii) Chaplygin gas models. Various DE models have been reviewed in the literatures including \cite{DE1,DE2}. It may be noted that around 68.3$\%$ of the total energy density of the present observable universe is contributed by DE. Remaining densities are due to dark matter (DM), ordinary baryonic matter and radiation. However, the contribution due to baryonic matter and radiation are negligible with respect to the total density of the universe.

Nojiri and Odintsov \cite{vis1,vis2} developed cosmological models, where the DE and DM were treated as imperfect fluids. Viscous fluids represent one particular case of what was presented in \cite{vis1,vis2}. In recent years, a handful of literatures have explored the possibility that the late-time acceleration is driven by a kind of viscous fluid \cite{vis3,vis4,vis5}. In those references \cite{vis3,vis4,vis5}, a universe filled with bulk-viscous matter has been analysed through the theory for evolution of the viscous pressure under the perview of late-time acceleration of the universe. At this juncture, it should be stated that the late-time accelerated phase is not the only accelerated phase of the universe. There was another phase of acceleration in the early stage of the universe and that is called an inflationary scenario \cite{vis5}. In this very early phase of evolution, the dissipative effects including both bulk and shear viscosity are thought to play a significant role \cite{vis6}. It was reported in the work of Chimento et al. \cite{vis7} that it is possible to have the accelerated expansion of the universe in presence of a combination of a cosmic fluid characterised by bulk-dissipative pressure and a quintessence matter. It was also reported in \cite{vis7} that the above process involves a sequence of dissipative processes. The introductory attempts in the direction of creating the theory of relativistic dissipative fluids where reported in the works of Eckart \cite{vis8} and Landau and Lipshitz \cite{vis9}. A time varying viscosity in DE framework was reported by Nojiri and Odintsov \cite{vis10}, where EoS was considered to be associated with an inhomogeneous and Hubble parameter dependent term. Brevik et al. \cite{vis4} demonstrated the entropy for a coupled fluid and established a relationship between the entropy of closed FRW universe to the energy contained in it. Brevik et al. \cite{vis5} considered a DE - DM interacting scenario in a flat FRW universe and demonstrated Little Rip, Pseudo Rip and Bounce Cosmology in Bulk Viscosity framework by considering the bulk-viscous pressure as a function of Hubble parameter. In a very recent work, \cite{vis11} proposed an approach where an extended cosmological model was demonstrated in the context of viscous DE. Recent work of viscous cosmology also studied in \cite{vis12, vis13}.

It has already been mentioned in the previous paragraph that one of the broad type of DE candidates is the holographic DE (HDE) model that has been extensively discussed in the references \cite{HDE1,HDE2,HDE3,infl4}. The HDE is based on the holographic principle \cite{HDE1}. The density $\rho_{\Lambda}$  of HDE is given by $\rho_{\Lambda}=3 c^2 {M_P}^2 L^{-2}$ \cite{HDE3} where $c^2$ represents a dimensionless constant, $M_P$ is the reduced Plank mass and $L$ stands for IR cutoff. Different modifications to the IR cutoff has been proposed in the literature and various types of HDE have been discussed till date. Examples include  modified HDE \cite{HDE5}, Holographic Ricci DE \cite{HDE6} and generalised HDE \cite{HDE7}. Note that all these HDEs are just particular cases of Nojiri-Odintsov cut-off which may even serve to get the covariant theory for specific Nojiri-Odintsov cut-off \cite{NO1}. Since DE is responsible for about 68.3$\%$ of the total energy density of the late-time universe and it was negligible after the big-bang, Chen and Jing \cite{HDE8} argued that the DE density should be a function of the Hubble parameter $H$ and its higher order derivatives with respect to cosmic time $t$. The physical explanation behind this argument is that the $H$ gives us information about the expansion rate of the universe. Based on this physical explanation,  reference \cite{HDE8} proposed the following form of HDE which is basically a specific case of the Nojiri - Odintsov  HDE  \cite{odi1}:
\begin{equation}\label{h}
\rho_{\Lambda} = 3 (\alpha \ddot{H} H^{-1} + \beta \dot{H}  + \epsilon H^2)
\end{equation}
where $\alpha$, $\beta$ and $\epsilon$ are three arbitrary dimensionless parameters. It may be noted that we have assumed ${M_P}^2 = (8 \pi G)^{-1} = 1$. In the limiting case with $\alpha = 0$, we get the HDE with Granda-Oliveros (GO) cutoff. A detailed account in this regard has been presented in a recent work \cite{HDE10}. Recently holographic bounce was proposed in \cite{HB1}.

In this paper, firstly will reconstruct Hubble parameter $H$ without any choice of scale factor, also reconstruct state equation $w_{\Lambda}$, $w_{eff}$ and deceleration parameter. Secondly, with power law form of scale factor we will reconstruct Hubble parameter $H$, bulk viscous pressure $\Pi$, state equation $w_{\Lambda}$, $w_{eff}$ and deceleration parameter $q$ and the state finder parameter $r$ and $s$. Thirdly, in case of non interacting scenario in presence of bulk viscosity we will discuss equation of state $w_\Lambda$ , $w_{eff}$ and squared speed of sound. In the fourth in case of interacting scenario in presence of bulk viscosity we will discuss equation of state $w_{\Lambda}$, $w_{eff}$ and squared speed of sound. In the fifth we will discuss background evolution. Here we discus viscous interacting dark energy as scalar field. We reconstructed here the Hubble slow roll parameters $\epsilon_H$, $\eta_H$ and potential slow roll parameters $\epsilon_V$, $\eta_V$. Then reconstructed EoS parameter $w_{\phi}$ and reconstructed dissipative coefficient $\Gamma$ and calculated $2V-{\dot{\phi}}^2$. Rest of the paper is organized as follows: In section I, we have reported reconstruction schemes for $\rho_{\Lambda}$ through reconstruction of $H$ in presence of bulk viscosity with as well as without any specific choice of scale-factor. Interacting as well as non-interacting scenarios taken into account. In section II, we have demonstrated the findings of background evolution in viscous interacting dark energy as scalar field. We have calculated Hubble slow roll parameters, tensor to scalar ratio and the effective EoS parameter. We have concluded in section III.

\section{Reconstruction schemes}
\subsection{Viscous scenario without dark matter}
\subsubsection{Without any choice of scale factor.}
In the present subsection we are going to demonstrate a reconstruction scheme for the DE presented in Eq.(\ref{h}) in presence of bulk-viscosity. That is in addition to the thermodynamic pressure a bulk-viscous pressure is to be considered as $\Pi=-3H\xi$
where
\begin{equation}\label{mm}
\xi=\xi_0 + \xi_1 H + \xi_2 (\dot{H}+ H^2)
\end{equation}
where $\xi_0$, $\xi_1$, $\xi_2$ are all positive constraints.
In presence of bulk-viscosity the Friedmann's equations are:
\begin{equation} \label{sde}
H^2 = \frac{1}{3}\rho_\Lambda
\end{equation}
\begin{equation}
6\frac{\ddot{a}}{a}=-(\rho_{\Lambda}+3(p_{\Lambda}+\Pi))
\end{equation}
We shall now demonstrate reconstruction scheme for $\rho_{\Lambda}$ in presence of bulk-viscosity as stated above.
Solving Eq.(\ref{h}) and Eq.(\ref{sde}), we have the following solution for the Hubble parameter
\begin{equation}\label{pl}
H=\frac{2\alpha}{{\beta} t + C_0}
\end{equation}
where it should be stated that we have chosen $\epsilon=1$ within the permissible range. A natural consequence of the reconstructed $H$ in the reconstructed scale factor, whose form comes out to be
\begin{equation} \label{ade}
a=\frac{(\beta t + C_0)^{2\frac{\alpha}{\beta}}}{(\beta t_0 + C_0)^2\frac{\alpha}{\beta}}
\end{equation}
Using,the reconstructed Hubble parameter, we can get the reconstructed DE density as
\begin{equation} \label{ude}
\rho_{\Lambda}=\frac{12\alpha^2}{(\beta t + C_0)^2}
\end{equation}
Also, the bulk-viscosity coefficient being dependent upon $H$ and $\dot{H}$, we can have the following reconstructed bulk-viscous pressure:
\begin{equation} \label{p}
\Pi= \frac{- 6 \alpha (( C_0 + t \beta)^2 \xi_0 + 2 \alpha (C_0 + t \beta)\xi_1 + 4 \alpha^2 \xi_2 - 2 \alpha \beta \xi_2)}{( C_0 + t \beta )^3}
\end{equation}

\begin{figure}
\includegraphics[height=2.8in]{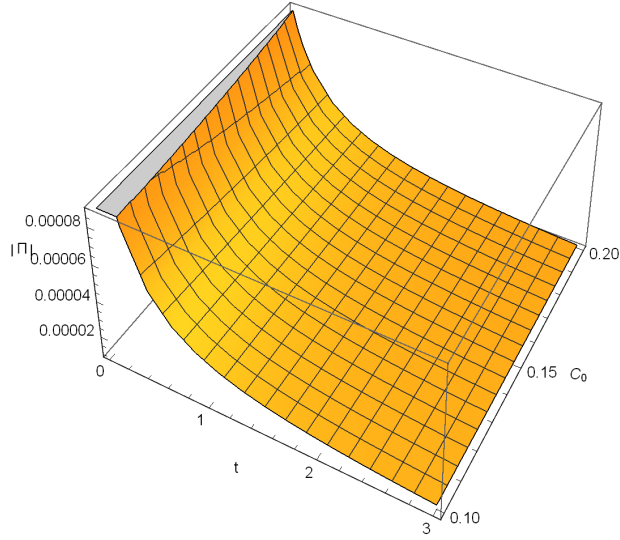}
\caption{Evolution of reconstructed Bulk Viscous pressure without any choice of scale factor. We consider $\alpha=0.00005$, $\beta = 0.53$, $\xi_0 = 0.06$, $\xi_1 = 0.023$, $\xi_2 = 0.00002$.}
\label{viss}
\end{figure}

As we are considering a D.E. dominated scenario under the assumption of negligible contribution due to DM, we are not supposed to have any interacting scenario. In absence of an interaction the conservation equation takes the following form in bulk-viscous framework:
\begin{equation} \label{jde}
\dot{\rho_{\Lambda}} + 3 H (\rho_{\Lambda} + p_{\Lambda} + \Pi) =0
\end{equation}
From Eq.(\ref{jde}), we can easily write $p_{\Lambda}=-(\frac{\dot{\rho_{\Lambda}}}{3H}+\rho_{\Lambda}+\Pi)$,which can be reconstructed using Equations (\ref{pl}), (\ref{ude}) and (\ref{p}).At this juncture,we have reconstructed $p_{\Lambda}$  and $\rho_{\Lambda}$ in bulk viscous scenario and hence we can have the EoS parameter $w_{\Lambda}=\frac{p_{\Lambda}}{\rho_{\Lambda}}$ in presence of bulk viscosity with background evolution as the holographic form of DE presented in Eq.(\ref{h}). The form of $w_{\Lambda}$ is derived below:
\begin{equation}
w_{\Lambda } = -1+\frac{2 \beta +3 \left(C_0+t \beta \right) \xi _0}{6 \alpha }+\xi _1 +\frac{(2 \alpha -\beta ) \xi _2}{C_0 +t
\beta }
\end{equation}
Clearly, the reconstructed behaves like $ w_\Lambda $ is quintessence, cosmological constant or phantom accordingly as $t \gtreqqless \frac{\frac{(\beta - 2 \alpha)\xi_2}{C_0} - \xi_1 - \frac{2 \beta + 3 C_0 \xi_0}{6\alpha}}{\frac{\beta \xi_0}{2 \alpha} - \frac{( 2 \alpha - \beta)\xi_2 \beta}{{C_0}^2}}$ and $ \frac{{C_0}^2 \xi_0}{\xi_2} \neq 4 \alpha^2 - 2 \alpha \beta$. Now, we consider the effective EoS parameter $w_{eff} = \frac{{p_{\Lambda}} + \Pi}{\rho_{\Lambda}}$, which comes out to be $w_{eff} = - 1 + \frac{\beta}{3\alpha}$ and the deceleration parameter becomes $q=-1+\frac{\beta}{2\alpha}$.

Hence, it is understandable that if $\frac{\beta}{3\alpha}>0$ then $w_{eff}>-1$ i.e., quintessence and similarly, for $\frac{\beta}{3\alpha}<0$ then $w_{eff}<-1$ i.e., phantom.We thus infer that although the presence of bulk-viscosity influences $w_{\Lambda}$, its impact is neutralised in $w_{eff}$. Hence, behaviour of $w_{eff}$ to be quintessence or phantom would be determined by the nature of $\alpha$ and $\beta$ as applied in Eq.(\ref{h}). Furthermore, accelerated expansion would be available if $q<0$ i.e., $-1+\frac{\beta}{2\alpha}<0$ i.e.,$\alpha>\frac{1}{2}\beta$. The behaviour of the bulk viscous pressure reconstructed in Eq.(\ref{p}) is plotted in Fig.\ref{viss}, where it is observed that in absence of DM the effect of bulk viscous pressure is decaying with cosmic time $t$.

 \subsubsection{With power law form of scale factor.}
 In the previous section, we have demonstrated the behaviour of bulk-viscous pressure, where the background evolution is according to the HDE type presented in Eq.(\ref{h}). In the previous section we didnot make any assumption regarding the choice of scale factor. Rather we have obtained the solution for the Hubble parameter to get the scale factor. In the present section we are going to develop a reconstruction scheme for the DE candidate presented in Eq.(\ref{h}) with power-law form of scale factor and in presence of bulk-viscosity. The scale factor is chosen as $ a = a_0 t^n$ where $a_0,n>0$. Therefore,
\begin{equation}\label{zuub}
H =\frac{n}{t}.
\end{equation}
 Hence, the D.E. density becomes
\begin{equation} \label{fce}
\rho_{\Lambda} = \frac{6 \alpha -3 n \beta +3 n^2 \epsilon }{t^2}
\end{equation}
and the bulk viscosity coefficient is
\begin{equation}
\xi =\frac{t^2 \xi_0 + n t \xi_1 +(-1+n) n \xi_2}{t^2}
\end{equation}
Finally, the bulk-viscous pressure $\Pi=-3H\xi$ comes out to be
\begin{equation} \label{pio}
\Pi = -\frac{3 n \left(t^2 \xi_0 +n t \xi_1 +(-1+n) n \xi_2\right)}{t^3}
\end{equation}
\begin{figure}
\includegraphics[height=2.8in]{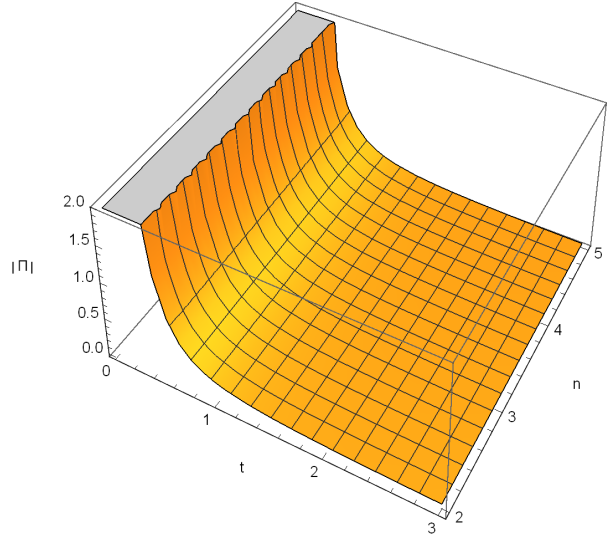}\\
\caption{Evolution of reconstructed Bulk Viscous pressure with power law form of scale factor. We consider $\xi_0 = 0.06$, $\xi_1 = 0.023$, $\xi_2 = 0.00002$.}
\label{bul}
\end{figure}
Using Eq.(\ref{fce}) in the conservation equation Eq.(\ref{jde}), we can find reconstructed $p_{\Lambda}$ in presence of bulk-viscosity and subsequently obtained the reconstructed EoS parameter as follows:
\begin{equation}\label{ooce}
w_{\Lambda} = \frac{3 n^2 t^2 \xi_0+t ((4-6 n) \alpha +n ((-2+3 n) \beta +n (2 \epsilon -3 n \epsilon +3 n \xi_1)))+3 (-1+n) n^3 \xi_2}{n t \left(6 \alpha -3 n \beta +3 n^2 \epsilon \right)}
\end{equation}.
Like the previous section,we obtained the EoS parameter as
\begin{equation}\label{zoob}
w_{eff}=-1+\frac{2}{3n}
\end{equation}
The deceleration parameter
\begin{equation}\label{zuin}
q=-1+\frac{1}{n}
\end{equation}
As $n$ is always positive, $w_{eff}>-1$ and hence it is always quintessence. Furthermore, in order the acceleration to exist we will $n>1$.

A geometrical diagnostic for DE model has been introduced by Sahni et al. \cite{DE3} and Alam et al. \cite{DE4}, which makes it possible to discriminate and classify various DE models. This helps us understand whether a specific model of DE is deviated from ${\Lambda}$CDM. The diagnostic pair is called as state finder pair and is denoted by $\{r,s\}$  where
\begin{equation}
r=\frac{\dddot{a}}{a H^3}
\end{equation}
\begin{equation}
s=\frac{r-1}{3(q-\frac{1}{2})}
\end{equation}
where $q=-\frac{a\ddot{a}}{{\dot{a}}^2}$. In the present case, we calculate the state finder parameters and observed that
\begin{equation}\label{rr}
r = \frac{(-2+n) (-1+n)}{n^2}
\end{equation}
\begin{equation}\label{ss}
s =\frac{2}{3 n}
\end{equation}
In order to have $r=1$, we need $n=\frac{2}{3}<1$, which is not compatible with the requirement of the present acceleration. Furthermore, $s\neq0$. Hence, for a fixed $n$ the ${\Lambda}CDM$ fixed $\{r=1,s=0\}$ is not attainable. However as $n\rightarrow\infty$ we observe that $r\rightarrow1,s\rightarrow0$. Therefore, we can understand that the $\Lambda$CDM fixed point is attainable in the limiting case. Moreover, it is also understand that like the previous case, the presence of bulk viscosity is not influencing the deceleration parameter and state-finder diagnostics.
\subsection{Viscous scenario in presence of dark matter}
\subsubsection{non-interacting scenario}
The conservation equation in a viscous scenario can be broken into two parts when we consider interaction between DE and DM. An interaction term $Q$ can be chosen in various forms and is added to the right hand side in the following manner:
\begin{equation}\label{oyo}
\dot{\rho_\Lambda} + 3H(\rho_\Lambda + p_\Lambda + \Pi)=-Q
\end{equation}
\begin{equation}\label{oyoo}
\dot{\rho_m} +3 H \rho_m =Q
\end{equation}
In this subsection we consider non interacting scenario i.e., $Q=0$ and hence from Eq.(\ref{oyoo}), we will have the solution for DM as $\rho_m=\rho_{m0}a^{-3}$. As we consider the coexistence of DE and DM, the first Friedmann equation takes the form $3H^2=\rho_m + \rho_{\Lambda}$ and hence using the Eq.(\ref{oyo}), we have [$\epsilon=1$],
\begin{equation} \label{reconstructed}
3 H^2 = 3 \left( \alpha \ddot{H} H^{-1} + \beta \dot{H} + H^2 \right) + \rho_{m0} a^{-3}
\end{equation}
where $ \dot{H} = a H \dfrac{dH}{da}$ and
      $\ddot{H} = aH \left( a\left(\dfrac{dH}{dt}   \right)^2 +     H\left(\dfrac{dH}{dt} + a \dfrac{d^{2} H}{d t^{2}}\right)\right)$ .
       Solving Eq.(\ref{reconstructed}) we have the reconstructed $H$ as a function of scale factor $a$ as follows:
\begin{equation} \label{positive}
H(a) = \pm \left( \frac{\sqrt{2}{\sqrt{{\frac{- \rho_{m0} \beta}{a^3}}+ 9 a^{- \frac{\beta}{\alpha}} \alpha(-3\alpha + \beta)C_1 + 9(3\alpha - \beta)\beta C_2}}}{3 \sqrt{(3 \alpha - \beta)\beta}}\right)
\end{equation}
  We consider the the positive solution in Eq.(\ref{positive}). Now we will try to constrain the model parameters present in the solution for $H$.
\begin{figure}
\includegraphics[height=2.8in]{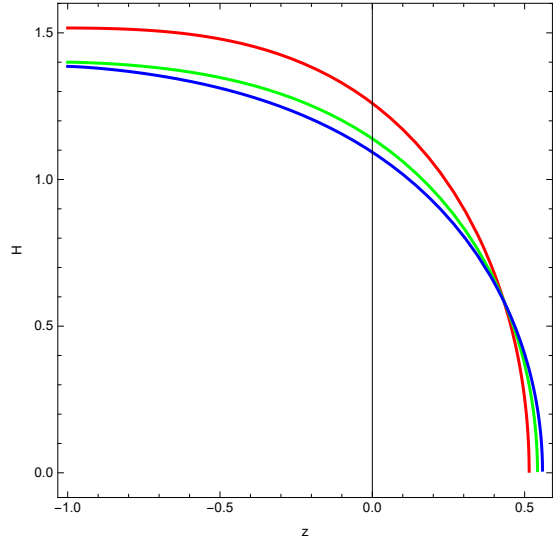}
\caption{ Evolution of reconstructed Hubble parameter in non interacting scenario. We have chosen $\rho_{m0} = 0.32$ . The red ,green and blue correspond to $\alpha = 0.09$, $\beta = 0.15$, $C_1=0.1$, $C_2 = 1.15$, $\eta=1.1$ ;$\alpha = 0.12$, $\beta = 0.18$, $C_1=0.2$, $C_2 = 0.98$, $\eta= 0.8$ and $\alpha = 0.15$, $\beta = 0.21$ , $C_1=0.3$ , $C_2 = 0.99$ , $\eta=0.7$ respectively.}
\label{recc}
\end{figure}
  Clearly if $\beta>0$ we must have the following in order to have a real solution for $H$.
\begin{equation}
\alpha > \frac{\beta}{3}
\end{equation}
Using the above constraint in the numerator we can further have
\begin{equation}\label{2}
C_2 > {\frac{1}{9 \beta}} \left(\frac{\beta \rho_{m0}}{3 \alpha - \beta} + 9 C_1 \alpha \right) a^{-\frac{\beta}{\alpha}}
\end{equation}
We can infer from (\ref{2}) that if $C_1 >0$ then $C_2 >0$. It is also understandable that it is feasible to choose $C_2 > C_1$. Hence we assume $C_2 = C_1 + \eta, ~~\eta >0$. This assumption leads us to have the following constraint:
\begin{equation} \label{C1}
C_1 > \frac{(\frac{1}{9 \beta}) (\frac{\beta \rho_{m0}}{3 \alpha - \beta}) a^{-\frac{\beta}{\alpha}} - \eta}{(1 - \frac{\alpha}{\beta} a^{- \frac{\beta}{\alpha}})}
\end{equation} \\
Using the positive solution of Eq.(\ref{positive}) and computing its derivative with respect to cosmic time $t$ as explained above we get the reconstructed DE density as
\begin{equation}\label{dens}
\rho_{\Lambda} = \frac{a^{-3-\frac{\beta }{\alpha }} \left(18 a^{3+\frac{\beta }{\alpha }} C_2 (3 \alpha -\beta ) \beta +18 a^3 C_1
\alpha  (-3 \alpha +\beta )+a^{\frac{\beta }{\alpha }} \rho_{m0} \beta  (-2-9 \alpha +3 \beta )\right)}{3 (3 \alpha -\beta ) \beta }
\end{equation}
Also, we can also reconstructed the viscosity as
\begin{equation}\label{xxx}
\begin{array}{c}
\xi = \xi _0+\frac{\sqrt{2} \sqrt{-\frac{\rho_{m0} \beta }{a^3}+9 {C_2} (3 \alpha -\beta ) \beta +9 a^{-\frac{\beta }{\alpha }}
{C_1} \alpha  (-3 \alpha +\beta )} \xi _1}{3 \sqrt{(3 \alpha -\beta ) \beta }}+\\
\frac{a^{-3-\frac{\beta }{\alpha }} \left(a^{\frac{\beta }{\alpha
}} \rho_{m0} \beta +18 a^{3+\frac{\beta }{\alpha }} {C_2} (3 \alpha -\beta ) \beta -9 a^3 {C_1} \left(6 \alpha ^2-5 \alpha  \beta +\beta
^2\right)\right) \xi _2}{9 (3 \alpha -\beta ) \beta }
\end{array}
\end{equation}
which helps us get the bulk-viscous pressure as
\begin{equation}\label{ppp}
\begin{array}{c}
\Pi = -\frac{\sqrt{2} \sqrt{-\frac{\rho_{m0} \beta }{a^3}+9 {C_2} (3 \alpha -\beta ) \beta +9 a^{-\frac{\beta }{\alpha }} {C_1}
\alpha  (-3 \alpha +\beta )} }{\sqrt{(3 \alpha -\beta ) \beta }}\times\\
\left(\xi _0+\frac{\sqrt{2} \sqrt{-\frac{\rho_{m0} \beta }{a^3}+9 {C_2} (3 \alpha -\beta ) \beta +9 a^{-\frac{\beta
}{\alpha }} {C_1} \alpha  (-3 \alpha +\beta )} \xi _1}{3 \sqrt{(3 \alpha -\beta ) \beta }}+\frac{a^{-3-\frac{\beta }{\alpha }} \left(a^{\frac{\beta
}{\alpha }} \rho_{m0} \beta +18 a^{3+\frac{\beta }{\alpha }} {C_2} (3 \alpha -\beta ) \beta -9 a^3 {C_1} \left(6 \alpha ^2-5 \alpha  \beta
+\beta ^2\right)\right) \xi _2}{9 (3 \alpha -\beta ) \beta }\right)
\end{array}
\end{equation}
The above equations gives us the bulk-viscous pressure when the background evolution is governed by the DE given in Eq.(\ref{h}). As we already have reconstructed $H$, $\rho_{\Lambda}$ and $\Pi$, we can get $p_{\Lambda}$ from the conservation equation (see Eq.(\ref{oyo}))taking $Q=0$ with that $p_{\Lambda}$ we can have the following EoS parameter:
\begin{equation}\label{www}
\begin{array}{c}
w_{\Lambda } = \left(a^{3+\frac{\beta }{\alpha }} (3 \alpha -\beta ) \beta  \left(-6 a^{-\frac{\beta }{\alpha }}C_1+
\frac{\rho_{m0} \left(-3+\frac{2}{-3
\alpha +\beta }\right)}{a^3}+\frac{18 a^{-\frac{\beta }{\alpha }}C_1 \alpha  (3 \alpha -\beta )+\frac{\rho_{m0} (2+9 \alpha -3 \beta )
\beta }{a^3}+18 C_2 \beta  (-3 \alpha +\beta )}{(3 \alpha -\beta ) \beta }+\right.\right.\\
\left.\left.\left.\frac{3 \sqrt{2} \sqrt{-\frac{\rho_{m0} \beta }{a^3}+9 C_2 (3 \alpha -\beta ) \beta +9 a^{-\frac{\beta }{\alpha }}
C_1 \alpha  (-3 \alpha +\beta )} \zeta}{\sqrt{(3
\alpha -\beta ) \beta }}\right)\right)\times\right.\\
\left(18 a^3 C_1 \alpha  (-3 \alpha +\beta )+a^{\frac{\beta }{\alpha }} \beta  \left(18 a^3 C_2 (3 \alpha -\beta )+\rho_{m0}
(-2-9 \alpha +3 \beta )\right)\right)^{-1}
\end{array}
\end{equation}
where,

$\zeta=\left(\xi _0+\frac{1}{9} \left(\frac{3 \sqrt{2} \sqrt{9 a^{-\frac{\beta }{\alpha }} C_1 \alpha  (-3
\alpha +\beta )+\beta  \left(-\frac{\rho_{m0}}{a^3}+27 C_2 \alpha -9 C_2 \beta \right)} \xi _1}{\sqrt{(3 \alpha -\beta ) \beta }}+\left(18
C_2+\frac{\rho_{m0}}{a^3 (3 \alpha -\beta )}+\frac{9 a^{-\frac{\beta }{\alpha }} C_1 (-2 \alpha +\beta )}{\beta }\right) \xi _2\right)\right)$
\begin{figure}
\includegraphics[height=2.8in]{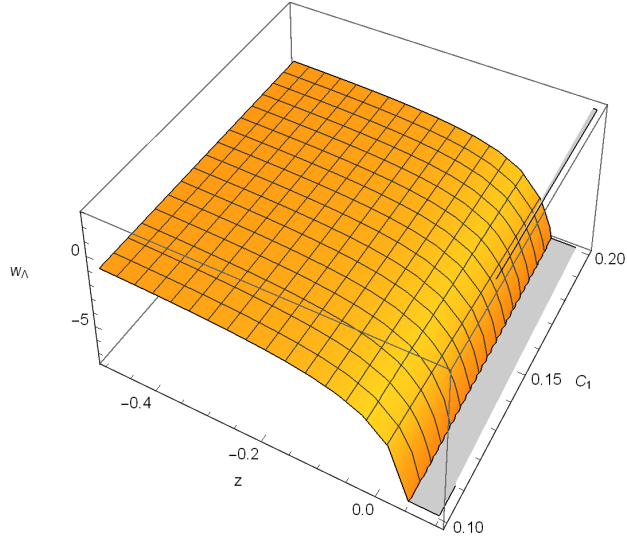}
\caption{Evolution reconstructed EoS parameter in non intracting scenario. We have chosen $\rho_{m0} = 0.32$, $\alpha=-0.12$, $\beta=-0.4$, $C_2=1.3$, $\xi_0 = 0.1203$, $\xi_1 =0.0105$, $\xi_2 = 0.003$.}
\label{bb}
\end{figure}

\begin{table}
\caption{Reconstructed EoS parameter for different combinations of {$\xi_0$, $\xi_1$, $\xi_2$} in the current universe (z=0) for non interacting viscous scenario}
\centering
\begin{tabular}{|c|c|c|c|}\hline
$z=0$ & $\xi_0=0.3$, $\xi_1=0.1332$, $\xi_2=0.0026$ & $\xi_0=0.3$, $\xi_1=0.122$, $\xi_2=0.0999$ & $\xi_0=0.5$, $\xi_1=0.0005$, $\xi_2=0.021$ \\ \hline
$w_{\Lambda}$ &{-1.08172} &{-0.904172} &{-0.995758} \\ \hline
\end{tabular}
 \label{tab1}
\end{table}

The Hubble parameter reconstructed in Eq.(\ref{positive}) is plotted in Fig.\ref{recc}, where against  redshift $z$ and we observed that $H>0$ and is having increasing pattern with evolution of the universe, which is consistent with the present accelerated expansion.

The squared speed of sound accounts for the speed of propagation of the perturbations of the energy density \cite{DEN1}, is now considered for the current model. The squared speed of sound is given by
\begin{equation}\label{vvv}
{v_s}^2 = \frac{\dot{p_{eff}}}{\dot{\rho_{\Lambda}}}
\end{equation}
This approach for checking stability of the DE model has earlier been used in Myung \cite{DE5}. The present
model is considering the presence of bulk-viscous pressure alongwith the thermodynamic pressure.
Hence, instead of considering $p_{\Lambda}$ only. We take to
compute ${v_s}^2$. Hence, we are going to consider ${v_s}^2 = \frac{\dot{p_{eff}}}{\dot{\rho_{\Lambda}}}$ and consequently the squared speed of sound comes out to be
\begin{equation}
{v_s}^2 = -\frac{2 a^3 C_1 (-3\alpha+\beta)^2}{\alpha \left(a^{\frac{\beta}{\alpha}} \rho_{m0}(2+9\alpha-3 \beta)+6 a^3 C_1(3 \alpha - \beta)\right)}
\end{equation}

\begin{figure}
\includegraphics[height=2.8in]{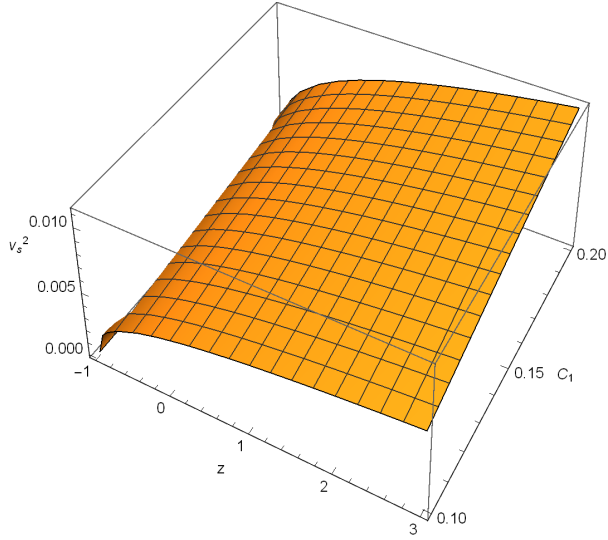}\\
\caption{ Evolution of reconstructed squared speed of sound in non-interacting scenario. We consider $\rho_{m0} = 0.32$, $\alpha= -0.12$, $\beta= -0.4$, $C_2 = 1.3$.}
\label{sss}
\end{figure}

Based on the constraints already obtained for the constants, the squared speed of sound is plotted in Fig.\ref{sss}, for a range of values of $C_1$ within its permissible boundaries. It has been observed that the squared speed of sound is positive throughout and for lower values of $C_1$ it is closed to zero, for the current universe i.e., $z=0$ and significantly greater than zero for higher values of $C_1$. Furthermore, it appears from Fig.\ref{sss} is also apparent that for higher value of $C_1$ the squared speed of sound will remain positive for a considerable period of time beyond $z=0$. Hence, a very stable model against small perturbations can be obtained from this non-interacting scenario where the background evolution is holographic and the universe is under bulk-viscous pressure apart from the thermodynamic pressure. If we consider the physical bounds $0\leq{v_s}^2\leq1$, it is apparent from Fig.\ref{sss} that this model is not violating the bounds.
\subsubsection{Interacting scenario in presence of bulk viscosity}
In the present section we are going to demonstrate the cosmological consequences of an interaction between the HDE with higher order derivatives as presented in Eq.(\ref{h})and the pressureless dark matter. The interaction term $Q$ is chosen in the form $Q=3H\delta \rho_m$ where $\delta$ is the interaction parameter and $\rho_m$ is the density of pressureless dark matter. The conservation equations in interacting scenario and in presence of bulk viscosity are
\begin{equation}\label{rts}
\dot{\rho_\Lambda} + 3H(\rho_\Lambda + p_\Lambda + \Pi)= -3H \delta \rho_m
\end{equation}
where $\Pi$, the bulk viscous pressure, has the form as descrbed in Eq.(\ref{mm})
\begin{equation}\label{kp}
\dot{\rho_m} +3 H \rho_m =3H \delta \rho_m
\end{equation}
As already mentioned previously, the First Friedmann equation in presence of DM takes the form $3 H^2 = \rho_m + \rho_{\Lambda} $, where $\rho_{\Lambda}$ comes from the solution of Eq.(\ref{kp})as $\rho_m = \rho_{m0} a^{-3(1-\delta)}$. With this form of DM, we obtained the solution for Hubble parameter from the Friedmann equations mentioned above:
\begin{equation}\label{adf}
H = \frac{\sqrt{-2 a^{-3+3 \delta } \rho_{m0} \beta +18 a^{-\frac{\beta }{\alpha }} (\beta +3 \alpha  (-1+\delta )) (-1+\delta) \left(-\alpha  {C_1}+a^{\frac{\beta }{\alpha }} \beta  {C_2}\right)}}{3 \sqrt{\beta  (\beta +3 \alpha  (-1+\delta )) (-1+\delta )}}
\end{equation}
\begin{figure}
\includegraphics[height=2.8in]{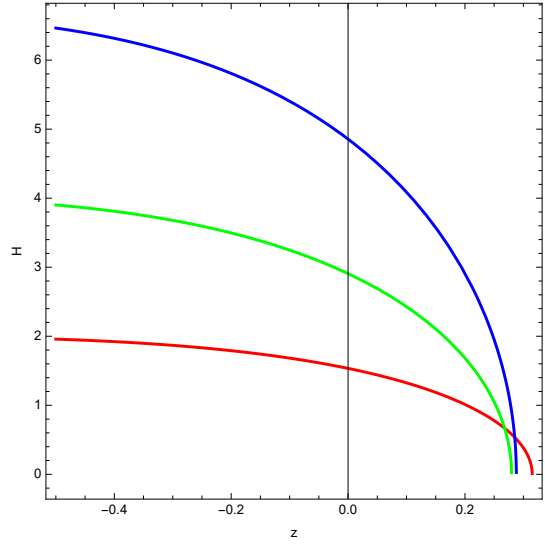}\\
\caption{Evolution of reconstructed Hubble parameter in interacting scenario. We have considered $\rho_{m0} = 0.32$,$\delta = 0.002$. The red, green and blue corresponds to $\alpha = 0.09$, $\beta= 0.25$, $C_1 = -2.8$, $C_2 = 2$;$\alpha = 0.09$, $\beta= 0.26$, $C_1 = 0.31$, $C_2 = 8.1$ and $\alpha = 0.09$, $\beta= 0.266$, $C_1 = 0.33$, $C_2 = 22.2$ respectively. }
\label{rq}
\end{figure}
It may be noted that like non-interacting case, here also we take $\epsilon=1$. As we know, $\dot{H}=a H \frac{dH}{da}$, we obtained the two derivatives of H below:
\begin{equation}
\dot{H}= \frac{a^{-3-\frac{\beta }{\alpha }} \left(-a^{\frac{\beta }{\alpha }+3 \delta } \rho_{m0}+3 a^3 {C_1} (\beta +3 \alpha (-1+\delta ))\right)}{3 (\beta +3 \alpha  (-1+\delta ))}
\end{equation}
\begin{equation}
\begin{array}{c}
\ddot{H} = -\frac{a^{-3-\frac{\beta }{\alpha }} \beta  \left(a^3 {C_1} \beta  (\beta +3 \alpha  (-1+\delta ))+a^{\frac{\beta }{\alpha}+3 \delta } \rho_{m0} \alpha  (-1+\delta )\right)}{3 \alpha  (\beta  (\beta +3\alpha  (-1+\delta )) (-1+\delta ))^{3/2}}\times\\\sqrt{-2 a^{-3+3 \delta } \rho_{m0} \beta +18 a^{-\frac{\beta }{\alpha }} \left(-{C_1}
\alpha +a^{\frac{\beta }{\alpha }} {C_2} \beta \right) (\beta +3 \alpha  (-1+\delta )) (-1+\delta )} (-1+\delta )
\end{array}
\end{equation}
As we are now having reconstructed Hubble parameter in interacting scenario, we apply this form Eq.(\ref{adf}) in Eq.(\ref{h}) to have the reconstructed HDE density and it comes out to be
\begin{equation}\label{re}
\begin{array}{c}
\rho_{\Lambda}=\frac{a^{-3-\frac{\beta }{\alpha }} \left(-a^{\frac{\beta }{\alpha }+3 \delta } \rho_{m0} \beta  \left(2+3 \beta  (-1+\delta )+9 \alpha  (-1+\delta )^2\right)-18 a^3 C_1 \alpha  (\beta +3 \alpha  (-1+\delta )) (-1+\delta )+18 a^{3+\frac{\beta }{\alpha }} C_2 \beta (\beta +3 \alpha  (-1+\delta )) (-1+\delta )\right)}{3 \beta  (\beta +3 \alpha  (-1+\delta )) (-1+\delta )}
\end{array}
\end{equation}
Furthermore, through this reconstructed $H$, we have the following forms of bulk viscosity coefficient and bulk viscous pressure respectively as
\begin{equation}\label{rc}
\begin{array}{c}
\xi = \xi _0+\frac{\sqrt{-2 a^{-3+3 \delta } \rho_{m0} \beta +18 a^{-\frac{\beta }{\alpha }} \left(-{C_1} \alpha +a^{\frac{\beta}{\alpha }} {C_2} \beta \right) (\beta +3 \alpha  (-1+\delta )) (-1+\delta )} \xi _1}{3 \sqrt{\beta  (\beta +3 \alpha  (-1+\delta )) (-1+\delta
)}} + \\  {\frac{a^{-3-\frac{\beta }{\alpha }} \left(-9 a^3 {C_1} (2 \alpha -\beta ) (\beta +3 \alpha  (-1+\delta )) (-1+\delta )+18 a^{3+\frac{\beta}{\alpha }} {C_2} \beta  (\beta +3 \alpha  (-1+\delta )) (-1+\delta )-a^{\frac{\beta }{\alpha }+3 \delta } \rho_{m0} \beta  (-1+3 \delta )\right)\xi _2}{9 \beta  (\beta +3 \alpha  (-1+\delta )) (-1+\delta )}}
\end{array}
\end{equation}
\begin{equation}\label{rw}
\begin{array}{c}
\Pi =-\frac{1}{\sqrt{\beta  (\beta +3 \alpha  (-1+\delta )) (-1+\delta )}}\sqrt{-2 a^{-3+3 \delta } \rho_{m0} \beta +18 a^{-\frac{\beta}{\alpha }} \left({C_1} \alpha +a^{\frac{\beta }{\alpha }} {C_2} \beta \right) (\beta +3 \alpha  (-1+\delta )) (-1+\delta )}\\{\left(\xi _0 +\frac{\sqrt{-2 a^{-3+3 \delta } \rho_{m0} \beta +18 a^{-\frac{\beta }{\alpha }} \left(-{C_1} \alpha +a^{\frac{\beta }{\alpha}} {C_2} \beta \right) (\beta +3 \alpha  (-1+\delta )) (-1+\delta )} \xi _1}{3 \sqrt{\beta  (\beta +3 \alpha  (-1+\delta )) (-1+\delta )}}+\right.} \\{\left.\frac{a^{-3-\frac{\beta }{\alpha }} \left(-9 a^3 {C_1} (2 \alpha -\beta ) (\beta +3 \alpha  (-1+\delta )) (-1+\delta )+18 a^{3+\frac{\beta}{\alpha }} {C_2} \beta  (\beta +3 \alpha  (-1+\delta )) (-1+\delta )-a^{\frac{\beta }{\alpha }+3 \delta } \rho_{m0} \beta  (-1+3 \delta )\right)\xi _2}{9 \beta  (\beta +3 \alpha  (-1+\delta )) (-1+\delta )}\right)}
\end{array}
\end{equation}
Since, $\rho_{\Lambda}$, $\rho_m$, $H$ and $\Pi$ are all having their reconstructed forms, we can reconstruct the thermodynamic pressure $p_{\Lambda}$ by putting the corresponding forms in Eq.(\ref{rts}) and hence the reconstructed EoS parameter $w=\frac{p_{\Lambda}}{\rho_{\Lambda}}$ comes out to be
\begin{equation}\label{rx}
\begin{array}{c}
w_{\Lambda}= {\left(3 a^{3+\frac{\beta }{\alpha }} \beta  (\beta +3 \alpha  (-1+\delta )) (-1+\delta ) \right.}\\
{\left(\frac{a^{-3-\frac{\beta }{\alpha }} \left(-6 a^3 \text{C1} (\beta +3 \alpha  (-1+\delta ))+a^{\frac{\beta }{\alpha }+3 \delta } \rho_{m0}
\left(2+3 \beta  (-1+\delta )+9 \alpha  (-1+\delta )^2\right)\right)}{3 (\beta +3 \alpha  (-1+\delta ))}-\right.}\\
{\frac{a^{-3-\frac{\beta }{\alpha }} \left(-a^{\frac{\beta }{\alpha }+3 \delta } \rho_{m0} \beta  \left(2+3 \beta  (-1+\delta )+9 \alpha
(-1+\delta )^2\right)-18 a^3 {C_1} \alpha  (\beta +3 \alpha  (-1+\delta )) (-1+\delta )+18 a^{3+\frac{\beta }{\alpha }} {C_2} \beta  (\beta
+3 \alpha  (-1+\delta )) (-1+\delta )\right)}{3 \beta  (\beta +3 \alpha  (-1+\delta )) (-1+\delta )}-}\\
{\frac{a^{-3 (1+\delta )} \rho_{m0} \sqrt{-2 a^{-3+3 \delta } \rho_{m0} \beta +18 a^{-\frac{\beta }{\alpha }} \left(-{C1} \alpha +a^{\frac{\beta
}{\alpha }} {C_2} \beta \right) (\beta +3 \alpha  (-1+\delta )) (-1+\delta )} \delta }{\sqrt{\beta  (\beta +3 \alpha  (-1+\delta )) (-1+\delta
)}}+}\\
{\frac{1}{\sqrt{\beta  (\beta +3 \alpha  (-1+\delta )) (-1+\delta )}}\sqrt{-2 a^{-3+3 \delta } \rho_{m0} \beta +18 a^{-\frac{\beta }{\alpha
}} \left(-{C_1} \alpha +a^{\frac{\beta }{\alpha }} {C_2} \beta \right) (\beta +3 \alpha  (-1+\delta )) (-1+\delta )} }\\
{\left(\xi _0+\frac{\sqrt{-2 a^{-3+3 \delta } \rho_{m0} \beta +18 a^{-\frac{\beta }{\alpha }} \left(-{C_1} \alpha +a^{\frac{\beta }{\alpha
}} {C_2} \beta \right) (\beta +3 \alpha  (-1+\delta )) (-1+\delta )} \xi _1}{3 \sqrt{\beta  (\beta +3 \alpha  (-1+\delta )) (-1+\delta )}}+\right.}\\
{\left.\left.\left.\left.\frac{a^{-3-\frac{\beta }{\alpha }} \left(-9 a^3 \text{C1} (2 \alpha -\beta ) (\beta +3 \alpha  (-1+\delta )) (-1+\delta
)+18 a^{3+\frac{\beta }{\alpha }} {C_2} \beta  (\beta +3 \alpha  (-1+\delta )) (-1+\delta )-a^{\frac{\beta }{\alpha }+3 \delta } \rho_{m0}
\beta  (-1+3 \delta )\right) \xi _2}{9 \beta  (\beta +3 \alpha  (-1+\delta )) (-1+\delta )}\right)\right)\right)\right.}\\
\left[\left(-a^{\frac{\beta }{\alpha }+3 \delta } \rho_{m0} \beta  \left(2+3 \beta  (-1+\delta )+9 \alpha  (-1+\delta )^2\right)-18 a^3 {C_1}
\alpha  (\beta +3 \alpha  (-1+\delta )) (-1+\delta )
\right.\right. \\ \left.\left.+18 a^{3+\frac{\beta }{\alpha }} {C_2} \beta  (\beta +3 \alpha  (-1+\delta )) (-1+\delta)\right)\right]^{-1}
\end{array}
\end{equation}

 \begin{figure}
\includegraphics[height=2.8in]{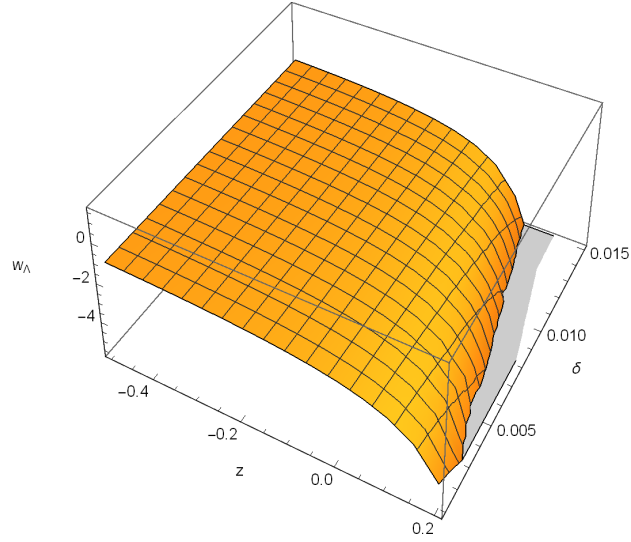}\\
\caption{ Evolution of reconstructed EoS parameter in interacting scenario. We consider $\alpha=0.09$ , $\beta= 0.25$ , $\rho_{m0} = 0.32$ ,$C_1 =-2.8$, $C_2 = 2$,$\xi_0=0.0012$, $\xi_1 = 0.007$, $\xi_2 =0.0001 $. }
\label{j}
\end{figure}

The reconstructed Hubble parameter and $w_{\Lambda}$ are now plotted in Fig.\ref{rq} and Fig.\ref{j} and Fig.\ref{j} respectively with choice of parameters in their acceptable ranges. In Fig.\ref{rq}, we observed that the Hubble parameter is increasing with the evolution of universe and in Fig.\ref{j}, we observe that the EoS parameter is tending to -1 and is behaving like phantom. Nevertheless it is not crossing the phantom boundary.
\begin{table}
\caption{Reconstructed EoS parameter for different combinations of {$\xi_0$, $\xi_1$, $\xi_2$} in the current universe (z=0) for interacting scenario}
\centering
\begin{tabular}{|c|c|c|c|}\hline
$z=0$ & $\xi_0=0.0012$, $\xi_1=0.8$, $\xi_2=0.0001$ & $\xi_0=1.2$, $\xi_1=0.0003$, $\xi_2=0.02$ & $\xi_0=1.4$, $\xi_1=0.00003$, $\xi_2=0.02$ \\ \hline
$w_{\Lambda}$ &{-1.00392} &{-0.955097} &{-0.81857} \\ \hline
\end{tabular}
\label{tab2}
\end{table}
As already discussed in the previous section, the squared speed of sound for the present case comes out to be
\begin{equation}\label{xk}
\begin{array}{c}
{v_s}^2 = {\left((\beta  (\beta +3 \alpha  (-1+\delta )) (-1+\delta ))^{3/2} \right.}\\
{\left(-2 \sqrt{2} a^6 {C_1} (3 \alpha -\beta ) (\beta +3 \alpha  (-1+\delta )) \sqrt{\Delta} \sqrt{\beta  (\beta +3 \alpha  (-1+\delta )) (-1+\delta )}+\right.}\\
{a^{3+\frac{\beta }{\alpha }+3 \delta } \rho_{m0} \alpha  \left(-54 {C_2} \beta  (\beta +3 \alpha  (-1+\delta ))^2 (-1+\delta )^2+\sqrt{2}
\sqrt{\Delta} \right.}
{\left.\left(2+3 \beta  (-1+\delta ) \right.\right.} \\ \left.\left. +9 \alpha  (-1+\delta )^2\right)
 \sqrt{\beta  (\beta +3 \alpha  (-1+\delta )) (-1+\delta )}\right) \delta
+9 a^{\frac{\beta }{\alpha }+6 \delta } \rho_{m0}^2 \alpha  \beta  (\beta +3 \alpha  (-1+\delta )) (-1+\delta ) \delta +\\
{\left.\left.\left.9 a^{3+3 \delta } {C_1} \rho_{m0} \alpha  (\beta +3 \alpha  (-1+\delta ))^2 (-\beta +6 \alpha  (-1+\delta )) (-1+\delta
) \delta \right)\right)\right.}\\
\left[{\left(a^3 \alpha  \beta ^2 \left(6 a^3 {C_1} (\beta +3 \alpha  (-1+\delta ))-a^{\frac{\beta }{\alpha }+3 \delta } \rho_{m0} \left(2+3
\beta  (-1+\delta )+9 \alpha  (-1+\delta )^2\right)\right) (\beta +3 \alpha  (-1+\delta ))^2
\right.}\right. \\ \left. \sqrt{-2 a^{-3+3 \delta } \rho_{m0} \beta +18 a^{-\frac{\beta
}{\alpha }} \left(-{C_1} \alpha +a^{\frac{\beta }{\alpha }} {C_2} \beta \right) (\beta +3 \alpha  (-1+\delta )) (-1+\delta )} (-1+\delta)^2\right)]^{-1}
\end{array}
\end{equation}
where\\
$\Delta={\sqrt{a^{-3-\frac{\beta }{\alpha }} \left(-a^{\frac{\beta }{\alpha }+3 \delta } \rho_{m0} \beta -9 a^3 {C_1} \alpha  (\beta
+3 \alpha  (-1+\delta )) (-1+\delta )+9 a^{3+\frac{\beta }{\alpha }} {C_2} \beta  (\beta +3 \alpha  (-1+\delta )) (-1+\delta )\right)}}$
Eq.(\ref{xk}) is plotted against redshift for a range of values of $\delta$. It is observed in this figure that the physical bounds $0\leq{v_s}^2\leq1$ are not violated. Hence, this model is stable against small perturbations.

\begin{figure}
\includegraphics[height=2.8in]{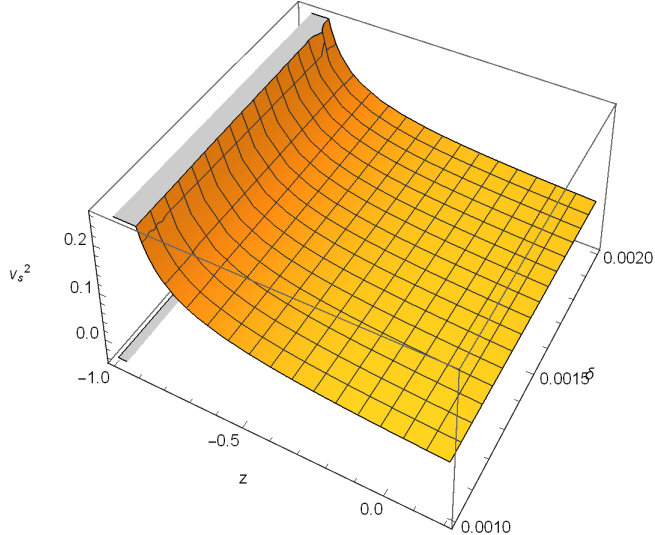}\\
\caption{ Evolution of reconstructed squared speed of sound for interacting viscous scenario. We considered $\alpha = 0.09$, $\beta= 0.25$, $\rho_{m0} = 0.32$,$C_2 = 2$.}
\label{km}
\end{figure}

\section{Viscous Interacting Dark Energy as Scalar Field.}

In the literature (e.g.\cite{infl1,infl2}), a number of inflationary self-interaction potentials have been proposed to date to explain the inflation. The self-interacting potentials result in different inflationary scenarios. In particular, if we talk about density fluctuations, then we will find that they have different observational consequences for the CMB radiation. In GR scalar field cosmology, different inflation potentials have been proposed in the literature \cite{infl3}. In a very recent work, Nojiri et al. \cite{odi1} applied the holographic principle at the early universe, obtained an inflation realization of holographic origin to calculate the Hubble slow-roll parameters and obtained the scalar spectral index, the tensor-to-scalar ratio, and the tensor spectral index. Bamba et al. showed the equivalence between scalar field theories and the fluid description \cite{DE1}. In the process firstly we take the described fluid and then derived a scalar field theory with the same EoS as that in a fluid description. By following this process, we got constraints on a coefficient function in the $\phi$ and $V(\phi)$ of the scalar field. Therefore ,derived the expression of $\phi$ and $V(\phi)$ in the scalar field theory for a fluid model. Again, we have a scalar field theory described by $\phi$ and $V(\phi)$ and by the solution of $\phi$, $V(\phi)$, $H$, we get $w_{\phi}$. By the expression of $\rho_{\Lambda}$ with $w_{\phi}$, we acquire $f(\rho)$ in the fluid description. Therefore, it implies that the scalar field theory and fluid model description is equivalent. In a a very significant work, \cite{infln1} demonstrated a phantom cosmology having a dynamics that allows the the universe to trace back the evolution to the inflationary epoch and developed the unified phantom cosmology where the same scalar field is capable of explaining the early time (phantom) inflation and late-time accelerated DE phase of the universe. Considering the inflationary dynamics in modified gravity framework of $f(R,G)$, the authors \cite{infln2} could demonstrate a double inflationary scenario. Inflationary dynamics through scalar field have also been demonstrated in \cite{infln3}, where inflationary solutions could be obtained that followed neither from any effective scalar field potential nor from a cosmological
constant.

In the present section we consider flat FRW universe to get the viscous interacting dark energy in scalar field framework. Denoting $\phi$ as a scalar field and $V(\phi)$ as a potential we have the following equations:
\begin{equation}\label{scs}
H^2 = \frac{1}{3} \left[\frac{1}{2} \dot{\phi}^2 + V(\phi)\right]
\end{equation}
\begin{equation}\label{sch}
\dot{H} = - \frac{1}{2} \dot{\phi}^2
\end{equation}
The equation of motion for the scalar field is
\begin{equation}\label{sca}
\ddot{\phi} + 3 H\dot{\phi}+\partial_\phi V(\phi) = 0
\end{equation}

The basic purpose for this section is to demonstrate whether it is possible to have inflationary expansion from the interacting viscous holographic dark energy in scalar field formalism. To do the same we consider slow roll parameters $\epsilon_H$ and $\eta_H$ given by the following equations:

In Eq. (\ref{sca}), if we consider standard approximation technique for analysing  inflation in slow roll approximation, then we have \cite{SRO}
$3 H \dot{\phi} \approx - \partial_\phi V(\phi)$
and from Eq. (\ref{scs}), we have $H^2 \approx \frac{1}{3} V(\phi)$.
For this approximation to be valid we need to have
\begin{equation}
\epsilon_H \ll 1,~~~~~~  \shortmid\eta_H(\phi)\shortmid\ll 1
\end{equation}
where $\epsilon_H$ and $\eta_H$  are given by
\begin{equation}\label{acb}
\frac{\ddot{a}}{a} = H^2 ( 1 - \epsilon_H )
\end{equation}
and
\begin{equation}\label{accbb}
\eta_H = \frac{- \ddot{\phi}}{H \dot{\phi}}
\end{equation}
These are Hubble slow roll parameters and in terms of potential, they can be written as []
\begin{equation}\label{agy}
\epsilon_V = \frac{1}{2} \left[\frac{\frac{dV}{d\phi}}{V}\right]^2
\end{equation}
\begin{equation}\label{aaggyy}
\eta_V= \frac{(\frac{dV}{d\phi})^2}{V}
\end{equation}
The slow roll parameters in term of potential become equal to the Hubble slow roll parameter in the following situation \cite{SROO}
\begin{equation}
\epsilon_V \approx \epsilon_H
\end{equation}
\begin{equation}
\eta_V \approx \epsilon_H + \eta_H
\end{equation}
It may be noted that $\epsilon_V$ is positive by definition and  slow-roll approximation is valid and inflation is guaranteed if the slow roll roll approximation i.e., $\epsilon_V\ll 1$ holds. In order to calculate $\epsilon_H$ as per Eq. (\ref{acb}), we need the scale factor and Hubble Parameter, which is already obtained in equation Eq. (\ref{adf}) for interacting DE in presence of bulk viscosity. The expression of $\epsilon_H$ for the scenario under consideration is presented below:
\begin{equation}\label{abc}
\epsilon_H=\frac{3 \beta  \left(a^{\frac{\beta }{\alpha }+3 \delta } \rho_{m0}-3 a^3 {C_1} (\beta +3 \alpha  (-1+\delta ))\right) (-1+\delta
)}{-2 a^{\frac{\beta }{\alpha }+3 \delta } \rho_{m0} \beta -18 a^3 \left({C_1} \alpha -a^{\frac{\beta }{\alpha }} {C_2} \beta \right)
(\beta +3 \alpha  (-1+\delta )) (-1+\delta )}
\end{equation}
Eq. (\ref{abc}) makes it apparent that the following constraints on $C_1$ as in Eq.(\ref{adf}) are to be satisfied:
For $\epsilon_H>0$ and $\beta>0$
\begin{equation}\label{a}
C_1>\frac{a^(\frac{\beta}{\alpha}+3\delta-3)\rho_{m0}}{3(\beta+3\alpha(-1+\delta))}
\end{equation}

For $\epsilon_H>0$ and $\beta<0$
\begin{equation}\label{b}
C_1<\frac{a^(\frac{\beta}{\alpha}+3\delta-3)\rho_{m0}}{3(\beta+3\alpha(-1+\delta))}
\end{equation}

For $\epsilon_H<1$
\begin{equation}\label{c}
C_1<\frac{18 a^{(\frac{\beta}{\alpha}+3)}C_2 \beta \Omega (-1+\delta)-2 a^{(\frac{\beta}{\alpha}+3\delta)}\rho_{m0}\beta-3\beta a^{(\frac{\beta}{\alpha}+3\delta)}\rho_{mo}(-1+\delta)}{9 a^3 \Omega (-1+\delta)[ 2 \alpha - \beta]}
\end{equation}
where $\Omega = \beta +3\alpha(-1+\delta)$

In the above constraints,(\ref{a}) and (\ref{b}) encompass the cases of positive as well as negative $\beta$. However both of them satisfy the positivity of $\epsilon_H$. The third inequation (\ref{c}) constraints $C_1$, which is not effected by the sign of $\beta$.

\begin{figure}
\includegraphics[height=2.8in]{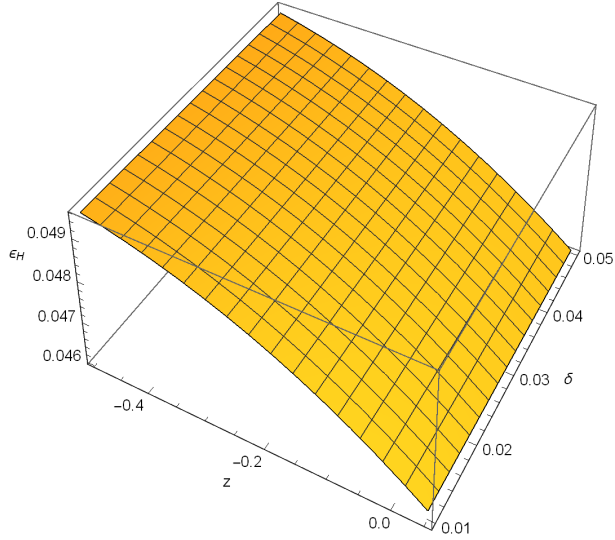}
\caption{ Evolution reconstructed of $\epsilon_H$ in case of viscous interacting DE as scalar field. We consider $\rho_{m0} = 0.32$, $\alpha= 0.09$, $\beta=0.009$, $C_1 = -6$, $C_2 = 0.0007$.}
\label{ep}
\end{figure}
The evolution of Hubble slow roll parameter $\epsilon_H$ is studied through Eq.(\ref{abc}) and Fig.\ref{ep}. Based on the constraints presented for $0<\epsilon_H<1$, the $\epsilon_H$ is plotted against the red shift $z$ for a range of values of $\delta$ in Fig.\ref{ep}. In this figure we observe that for the entire range of values of $\delta$, $\epsilon_H \ll 1$ and is always positive. Hence it is clear that under these constraints $\epsilon_H$ is an infinitesimally small positive number. Hence, we may look into the primordial inflation using this $\epsilon_H$ by describing the primordial inflation through quasi - de Sitter geometry with the EoS $w = -1 + \epsilon_H$. Clearly, because of the infinitesimally small $\epsilon_H$ we have $w\approx -1$. Therefore, it is possible to infer trivially that the effect of bulk-viscosity is not dominant during the early universe and during this phase the thermodynamic pressure will dominate and make the equation of state close to $-1$. The scalar field $\phi$ and potential $V(\phi)$ are now expressed in terms of scale factor $a$ as follows:
\begin{equation}
\dot{\phi}=\sqrt{-2 a^{-\frac{\beta }{\alpha }}{C_1}+\frac{2 a^{-3+3 \delta } \rho_{m0}}{3 (\beta +3 \alpha  (-1+\delta))}}
\end{equation}

\begin{equation}
V(\phi)=6 {C_2}+a^{-\frac{\beta }{\alpha }} \left({C_1}-\frac{6 {C_1} \alpha }{\beta }\right)-\frac{a^{-3+3 \delta } \rho_{m0}
(1+\delta )}{3 (\beta +3 \alpha  (-1+\delta )) (-1+\delta )}
\end{equation}
Using the solution of Hubble parameter as in Eq.(\ref{adf}) for interacting viscous scenario in Eq.(\ref{accbb}), we obtain the slow roll parameter as a function of scale factor as follows:
\begin{equation}\label{ett}
\eta_H=\frac{3 a^3 {C_1} \beta  (\beta +3 \alpha  (-1+\delta ))+3 a^{\frac{\beta }{\alpha }+3 \delta } \rho_{m0} \alpha  (-1+\delta
)}{-2 a^{\frac{\beta }{\alpha }+3 \delta } \rho_{m0} \alpha +6 a^3 {C_1} \alpha  (\beta +3 \alpha  (-1+\delta ))}
\end{equation}

\begin{figure}
\includegraphics[height=2.8in]{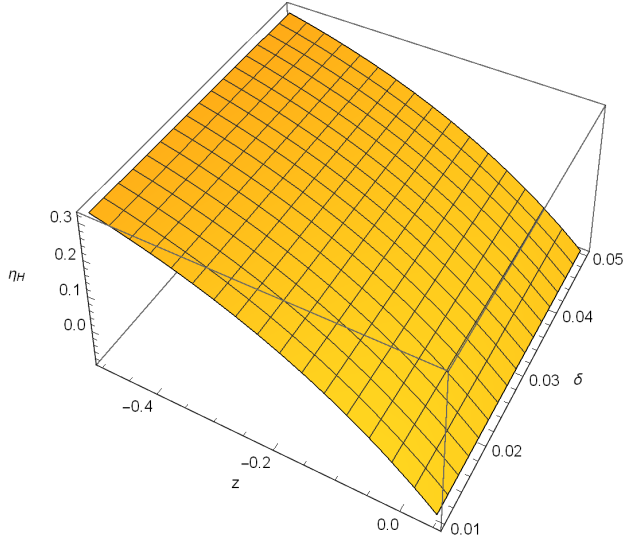}
\caption{ Evolution reconstructed  $\eta_H$ in case of viscous interacting DE as scalar field. We consider $\rho_{m0}= 0.32$, $\alpha=0.09$, $\beta=0.07$, $C_1 = -2$.}
\label{ettp}
\end{figure}
From the Fig.\ref{ep} and Fig.\ref{ettp}, we observe that $\eta_H$ and $\epsilon_H$ both are increasing. Therefore,the model has the scope of exit from inflation. Now we consider Eqns. (\ref{agy}) and (\ref{aaggyy}) to obtain the potential slow roll parameters in an interacting viscous scenario, where $H$ has been reconstructed in Eq.(\ref{adf}):
\begin{equation}\label{epv}
\begin{array}{c}
\epsilon_V=a^{-6-\frac{2 \beta }{\alpha }} \left(-2 a^{-3+3 \delta } \rho_{m0} \beta +18 a^{-\frac{\beta }{\alpha }} \left(-{C_1}
\alpha +a^{\frac{\beta }{\alpha }} {C_2} \beta \right) (\beta +3 \alpha  (-1+\delta )) (-1+\delta )\right)\\ \left(a^3 {C_1} (6 \alpha -\beta
) (\beta +3 \alpha  (-1+\delta ))-a^{\frac{\beta }{\alpha }+3 \delta } \rho_{m0} \alpha  (1+\delta )\right)^2\times\\ \left(18 \alpha ^2 \beta  \left(-2 a^{-\frac{\beta}{\alpha }} {C_1}+\frac{2 a^{-3+3 \delta } \rho_{m0}}{3 (\beta +3 \alpha  (-1+\delta ))}\right) (\beta +3 \alpha  (-1+\delta ))^3 (-1+\delta)\right.\\ \left. \left(6 {C_2}+a^{-\frac{\beta }{\alpha }} \left({C_1}-\frac{6 {C_1} \alpha }{\beta }\right)-\frac{a^{-3+3 \delta } \rho_{m0} (1+\delta)}{3 (\beta +3 \alpha (-1+\delta )) (-1+\delta )}\right)^2\right)^{-1}
\end{array}
\end{equation}

\begin{equation}\label{etv}
\begin{array}{c}
\eta_V={a^{-6-\frac{2 \beta }{\alpha }} \left(-2 a^{-3+3 \delta } \rho_{m0} \beta +18 a^{-\frac{\beta }{\alpha }} \left(-{C_1}\alpha +a^{\frac{\beta }{\alpha }} {C_2} \beta \right) (\beta +3 \alpha  (-1+\delta )) (-1+\delta )\right)}\\ \left(a^3 {C_1} (6 \alpha -\beta)(\beta +3 \alpha (-1+\delta ))-a^{\frac{\beta }{\alpha }+3 \delta} \rho_{m0} \alpha (1+\delta )\right)^2\times\\\left(9 \alpha ^2 \beta  \left(-2 a^{-\frac{\beta}{\alpha }} {C_1}+\frac{2 a^{-3+3 \delta } \rho_{m0}}{3 (\beta +3 \alpha  (-1+\delta ))}\right) (\beta +3 \alpha  (-1+\delta ))^3 (-1+\delta) \left(6 {C_2}+a^{-\frac{\beta }{\alpha }} \left({C_1}-\frac{6 {C_1} \alpha }{\beta }\right)\right.\right.\\ \left.\left.-\frac{a^{-3+3 \delta } \rho_{m0} (1+\delta
)}{3 (\beta +3 \alpha  (-1+\delta )) (-1+\delta )}\right)\right)^{-1}
\end{array}
\end{equation}
Relations (\ref{a})- (\ref{c}),we have derived conditions for $0< \epsilon_H <1$.Here, we further mention that for a inflation to occur one requires, $\eta_H\ll1$. Hence, using Eq.(\ref{ett}) we can constrain $C_1$ as follows:
\begin{equation}
C_1 \ll \frac{a^(\frac{\beta}{\alpha}+3\delta-3)\rho_{mo} \alpha(3\delta - 1)}{3 (2\alpha - \beta)(\beta + 3 \alpha (-1 + \delta))}
\end{equation}
The presence of promordial tensor fluctuations is predicted by many inflationary models. As with scalar fluctuations, tensor fluctuations are expected to follow a power law and are parametrised by the tensor index. We consider the tensor to scalar ratio \cite{TS1}
Tensor to scalar ratio:
\begin{equation}\label{ts}
r = \frac{16 \epsilon [1 - (C_E + 1)\epsilon]^2}{[ 1 - ( 2 C_E + 1)\epsilon + C_E \eta H]^2}
\end{equation}
where,$ C_E = 4 [ ln 2 + \gamma_E]-5$, $\gamma_E = 0.5772$ and $\eta$ depends upon second derivative of the potential and is having the form  $\eta = \frac{V^{''}(\phi)}{V(\phi)} - \frac{1}{2} \left[\frac{V^{'}(\phi)}{V(\phi)}\right]^2$. The scalar spectral index can now be expressed in terms of other slow roll parameters: $\epsilon = \frac{- \dot{H}}{H^2}$ , $\psi = \epsilon-\frac{\dot{\epsilon}}{2 H \epsilon}$ as follows:
\begin{equation}\label{nss}
n_s = 1 - 6 \epsilon + 2 \psi
\end{equation}
\begin{figure}
\includegraphics[height=2.8in]{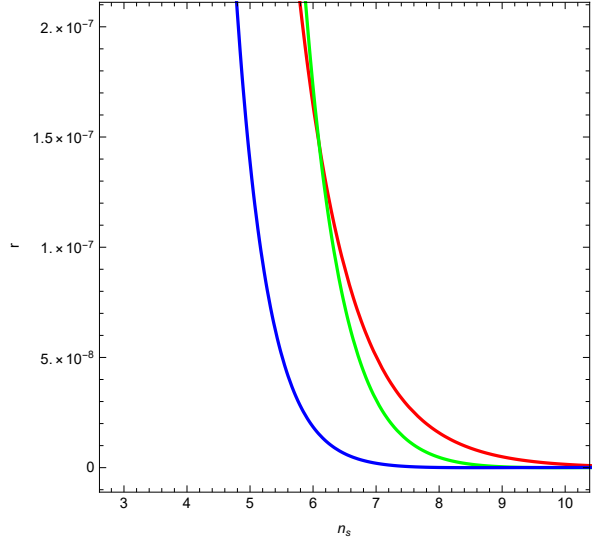}
\caption{ Evolution of reconstructed tensor-to-scalar ratio $r$ in case of viscous interacting DE as scalar field. We have considered $\rho_{m0} = 0.32$, $ \gamma_E = 0.5772 $, $\delta = 0.002$. The red, green and blue correspond to $\alpha= - 0.3$, $\beta=0.41$, $C_1=0.09$, $C_2=-0.0001$; $\alpha=-0.6$, $\beta=0.51$, $C_1= 0.08$, $C_2=-0.0002$ and $\alpha= -0.9$,$\beta= 0.61$, $C_1= 0.05$, $C_2=-0.0008$ respectively.}
\label{r}
\end{figure}
In Fig. \ref{r} we have plotted the tensor-to-scalar (Eq.(\ref{ts}))ratio predicted by this model as a function of scalar-spectral index $n_s$ (Eq.(\ref{nss})). It may be noted that the spectral index $n_s$ is obtained by using the reconstructed $H$ in the slow roll parameters $\epsilon$ and $\psi$. It is observed that the trajectories in the $n_s - r$ plane exhibit a decreasing behaviour, which is consistent with the observation of Jawad et al.\cite{jawad1}. It is also observed that $r<0.168$ (95$\%$ CL, Plank TT + LowP ) the observational bound found by Plank. Hence, the tensor-to-scalar ratio for this model is consistent with the observational bound due to Plank.  Hence this model can explain the primordial fluctuation in the early universe.

Now, we attempt to examine the availability of quasi exponential expansion for the viscous interacting scenario under consideration. In view of that we compute the effective EoS parameter for inflation and also investigate the behaviour of $2V-{\dot{\phi}}^2$. Using the equation of motion, the slow roll parameters can be written as \cite{SROOO}.

\begin{equation}
\epsilon = \frac{3}{2} {\dot{\phi}}^2 [\frac{1}{2} {\dot{\phi}}^2 + V(\phi)]^{-1}
\end{equation}

\begin{equation}
\psi = \frac{- \ddot{\phi}}{H \dot{\phi}}
\end{equation}

Also , the effective EoS parameter for the inflation is \cite{SROOOO}
\begin{equation}\label{ggc}
w_{\phi} = -1 + \frac{2}{3} \epsilon
\end{equation}
Using the expressions of $V(\phi)$ and $\dot{\phi}$ in Eq.(\ref{ggc}),we obtained the expression of $w_{\phi}$ as follows:
\begin{equation}\label{wph}
w_{\phi} = -1-\frac{\beta  \left(a^{\frac{\beta }{\alpha }+3 \delta } \rho_{m0}-3 a^3 {C_1} (\beta +3 \alpha  (-1+\delta))\right) (-1+\delta )}{a^{\frac{\beta }{\alpha }+3 \delta } \rho_{m0} \beta +9 a^3 \left({C_1} \alpha -a^{\frac{\beta }{\alpha }} {C_2}
\beta \right) (\beta +3 \alpha  (-1+\delta )) (-1+\delta )}
\end{equation}
\begin{figure}
\includegraphics[height=2.8in]{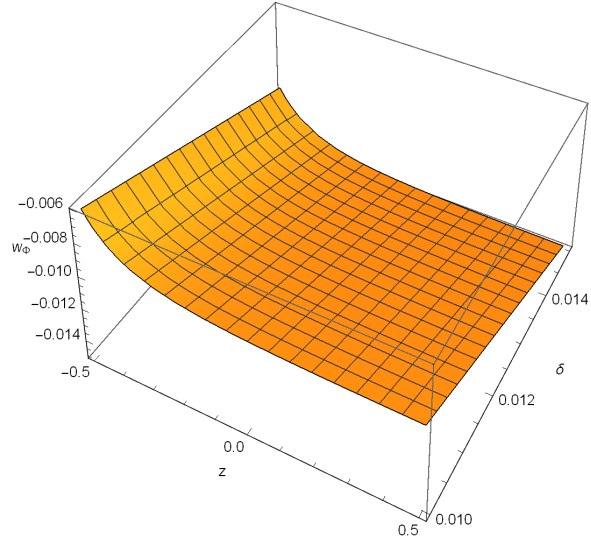}
\caption{ Evolution of reconstructed effective EoS parameter in case of viscous interacting DE as scalar field. We have considered $\rho_{m0}=0.32$ , $\alpha = 0.0019$, $\beta=-0.002$, $C_1  = 0.0008$, $C_2 = 0.0002$.}
\label{wphh}
\end{figure}
and obtained $2V-{\dot{\phi}}^2$
\begin{equation}\label{pott}
\begin{array}{c}
2 V - {\dot{\phi}}^2 = -2 a^{-3-\frac{\beta }{\alpha }} {C_1}^2 \left(3 a^3 {C_1} (3 \alpha -\beta ) (\beta +3 \alpha  (-1+\delta )) (-1+\delta
)\right. \\ \left. -9 a^{3+\frac{\beta }{\alpha }} {C_2} \beta  (\beta + 3 \alpha  (-1+\delta )) (-1+\delta )+ \right. \\ \left. a^{\frac{\beta }{\alpha }+3 \delta } \rho_{m0}
\beta  \delta \right)\times \left(\beta  (-1+\delta )\right)^{-1}
\end{array}
\end{equation}
\begin{figure}
\includegraphics[height=2.8in]{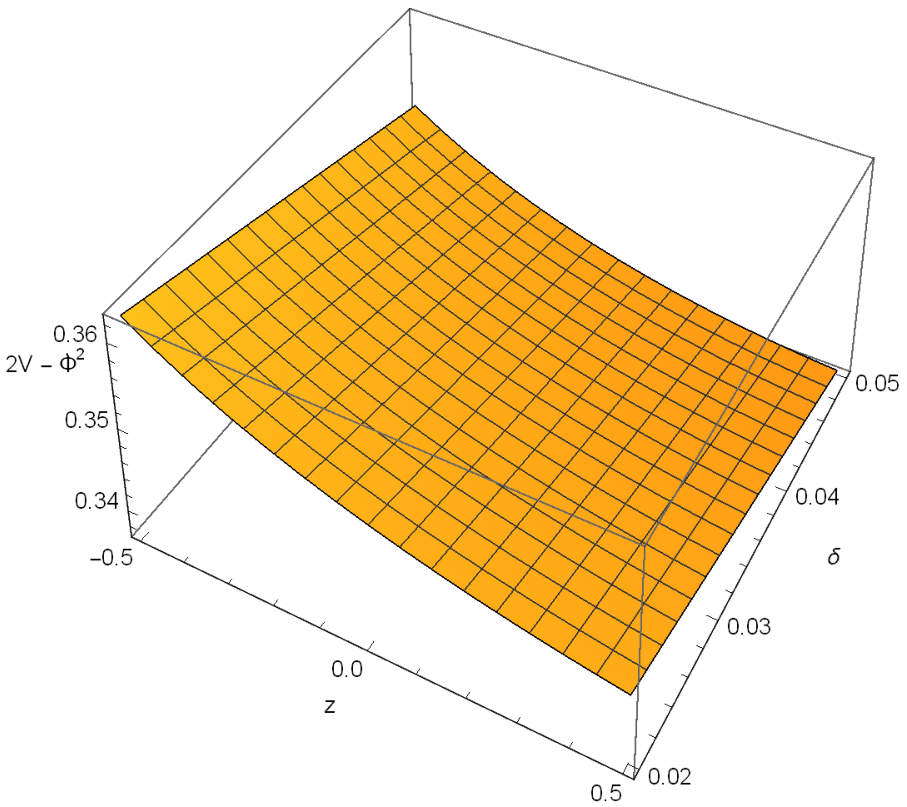}
\caption{ Evolution of reconstructed $2 v - {\dot{\phi}}^2$ in case of viscous interacting DE as scalar field. We consider $\rho_{m0}=0.32$, $\alpha=-0.04$, $\beta=0.002$,$C_1 = 0.2$, $C_2 = 0.001$.}
\label{pot}
\end{figure}

Therefore, $2 V - {\dot{\phi}}^2 > 0 $ for the range of $\delta$ in the range $0.02 \leq \delta \leq 0.05$

For the entire range,  $2 V - {\dot{\phi}}^2 > 0$, the difference goes higher with lowering in the value of $\delta$.
Hence, it can be interpreted that quasi exponential expansion is available for the interacting viscous holographic dark energy with higher order derivative. Furthermore, it is also observed that increase in the strength of interaction lowers the rate of expansion in presence of bulk viscosity.

\begin{equation}
\ddot{\phi} + ( 3 H + \Gamma )\dot{\phi} + \frac{dV}{d{\phi}} = 0
\end{equation}

\begin{figure}
\includegraphics[height=2.8in]{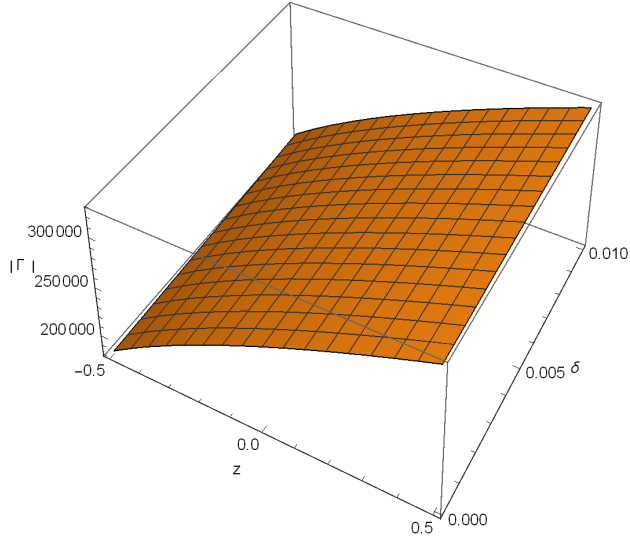}
\caption{ Evolution of reconstructed $\Gamma$ in case of viscous interacting DE as scalar field. We consider $\rho_{m0}=0.32$, $\alpha=-0.0006$, $\beta = 0.0000099$, $C_1 = 0.0004$, $C_2 = -0.07$.}
\label{gamee}
\end{figure}

\begin{figure}
\includegraphics[height=2.8in]{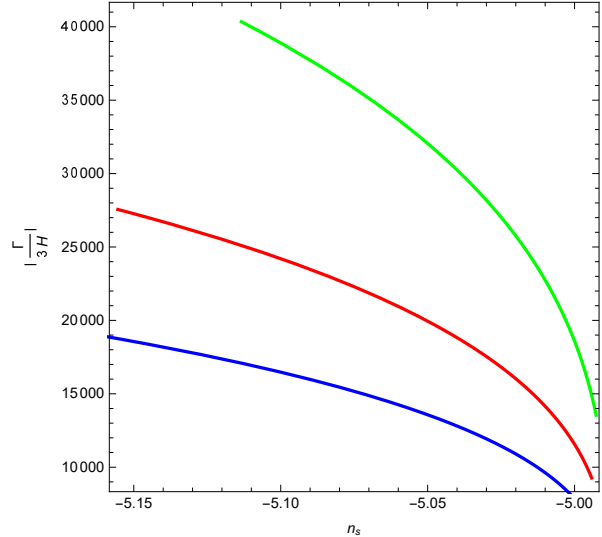}
\caption{ Evolution of reconsructed $\mid\frac{\Gamma}{3 H}\mid$ in case of viscous interacting DE as scalar field. We consider $\rho_{m0}=0.32$, $\delta =0.002$.For red , green and blue we consider  $\alpha=-0.0006$, $\beta = 0.0000099$, $C_1 = 0.0004$, $C_2 = -0.07$; $\alpha=-0.0005$, $\beta = 0.000008$,$C_1 = 0.0003$, $C_2 = -0.06$ and    $\alpha = -0.0008$, $\beta = 0.0000092$, $C_1 = 0.0005$, $C_2 = -0.08$  respectively.}
\label{games}
\end{figure}

$\Gamma$ represents the inflation decay rate or dissipative coefficient which is responsible for the decay of the scalar field into radiation during the inflationary expansion. We calculate $\Gamma$ from the following equation:
\begin{equation}\label{game}
\begin{array}{c}
\Gamma = \left[a^{\frac{1}{2} \left(3+\frac{\beta }{\alpha }\right)} \left(-a^{-\frac{3 \alpha +\beta }{2 \alpha }} {C_1} \left(a^3
{C_1} \beta  (\beta +3 \alpha  (-1+\delta ))+a^{\frac{\beta }{\alpha }+3 \delta } \rho_{m0} \alpha  (-1+\delta )\right) \right.\right.\\\left. \left(\sqrt{-2 a^{-3+3 \delta
} \rho_{m0} \beta +18 a^{-\frac{\beta }{\alpha }} \left(-{C_1} \alpha +a^{\frac{\beta }{\alpha }} {C_2} \beta \right) (\beta +3 \alpha
 (-1+\delta )) (-1+\delta )}\right)\right. \\ \left. \left(2 \alpha  \sqrt{a^{\frac{\beta }{\alpha }+3 \delta } \rho_{m0}-3 a^3 {C_1} (\beta +3 \alpha  (-1+\delta ))}
\sqrt{\beta  (\beta +3 \alpha  (-1+\delta )) (-1+\delta )}\right)^{-1}+ \right. \\ \left. \left( a \left(-2 a^{-3+3 \delta } \rho_{m0} \beta +18 a^{-\frac{\beta }{\alpha }}
\left(-{C_1}
\alpha +a^{\frac{\beta }
{\alpha }} {C_2} \beta \right)
(\beta +3 \alpha  (-1+\delta )) (-1+\delta )\right)\right.\right.\\\left.\left. \left(-a^3 {C_1}
(6 \alpha -\beta ) (\beta +3 \alpha  (-1+\delta ))+a^{\frac{\beta }{\alpha }+3 \delta } \rho_{m0} \alpha  (1+\delta )\right)\right) \right. \\ \left. \left({6 \alpha  \beta
 \left(a^{\frac{\beta }{\alpha }+3 \delta } \rho_{m0}-3 a^3 {C_1} (\beta +3 \alpha  (-1+\delta ))\right) (\beta +3 \alpha  (-1+\delta )) (-1+\delta
)}\right)^{-1}\right] \times \\ \left. \left({{C_1} \sqrt{a^{\frac{\beta }{\alpha }+3 \delta } \rho_{m0}-3 a^3 {C_1} (\beta +3 \alpha  (-1+\delta ))}}\right)^{-1}\right.
\end{array}
\end{equation}
Evolution of $\Gamma$ is observed in Fig.\ref{gamee}. We observed that $\mid \Gamma \mid > 1$, this implies that decay of scalar field into radiation during inflationary phase i.e., warm inflation.
$\mid \frac{\Gamma}{3 H}\mid$ measures the relative strength of thermal damping compared to an expansion damping. In warm inflation, $\mid \frac{\Gamma}{3 H}\mid<<1$ is the weak dissipative regime and Hubble damping is still the dominant term and if $\mid \frac{\Gamma}{3 H}\mid\geq1$, then it is strong dissipative regime. The evolution of $\mid \frac{\Gamma}{3 H}\mid$ aginst $n_s$ is observed in Fig.\ref{games}. We observe that $\mid \frac{\Gamma}{3 H}\mid\geq$1, hence it is a strong dissipative regime.

\section{Concluding remarks}
In the work reported above, we aimed at reconstructing $\rho_{\Lambda}$ through $H$ in non-interacting and interacting scenario and holographic background evolution. The bulk viscous pressure has been taken as $\Pi=-3H\xi$, where $\xi=\xi_0+\xi_1 H + \xi_2 (\dot{H} + H^2)$. In the reconstruction scheme reported here, firstly we choose viscous scenario neglecting the contribution of dark matter and without any choice of scale-factor. Considering the DE density mentioned in Eq.(\ref{h}), we have reconstructed $H$ through the First Friedmann's equation. As $H=\frac{\dot{a}}{a}$, we derived a solution for the scale factor (see Eq.(\ref{ade}))and also reconstructed DE density (see Eq.(\ref{ude})). We then got reconstructed bulk viscous pressure $\Pi$ in Eq. \ref{p}  and then  plotted $\mid\Pi\mid$ (see Fig. \ref{viss}) and found that the effect of bulk viscosity is decreasing with expansion of the universe. We then got EoS and from there we can see $w_{\Lambda}$ is quintessence,cosmological constant or phantom accordingly as $t \lesseqgtr\frac{\frac{(\beta - 2 \alpha)\xi_2)}{C_0} - \xi_1 - \frac{(2 \beta + 3 C_0 \xi_0)}{6 \alpha}}{\frac{\beta \xi_0}{2 \alpha} - \frac{(2 \alpha - \beta) \xi_2 \beta)}{{C_0}^2}}$ and $\frac{{C_0}^2 \xi_0}{\xi_2} \neq 4 {\alpha}^2 - 2 \alpha \beta$ and we got deceleration parameter as $q=-1+\frac{\beta}{2\alpha}$. If $\frac{\beta}{3 a}>0$ then it is quintessence and if $\frac{\beta}{3 a}<0$, then it is phantom.

Next, we choose viscous scenario neglecting the contribution of dark matter and with choice of scale-factor, then we got reconstructed Hubble parameter $H$ (see Eq.(\ref{zuub})), Bulk viscous pressure $\Pi$ (see Eq.(\ref{pio})), density $\rho_{\Lambda}$ (see Eq. (\ref{fce})), $w_{\Lambda}$ (see Eq.(\ref{ooce})) and $w_{eff}$ (see Eq. (\ref{zoob})). Plotted $\mid\Pi\mid$ (see Fig.\ref{bul}) and found that the effect of bulk viscosity is decreasing with expansion of the universe. We also derived deceleration parameter, $q=-1+\frac{1}{n}$ (see Eq.(\ref{zuin})) and the state finder parameter: $r=\frac{(-2+n)(-1+n)}{n^2}$ (see Eq. (\ref{rr})) and $s=\frac{2}{3n}$ (see Eq. (\ref{ss})). As $n$ is always positive it implies that $w_{eff} > -1$, so it is always quintessence. If $n>1$ then acceleration exists.

Next, we studied presence of dark matter in non-interacting scenario in presence of bulk-viscosity. So, we take $\delta=0$ and $\rho_m=\rho_{m0}a^{-3}$ and we get reconstructed $H$ (see Eq.(\ref{positive})) and then  plotted graph of the reconstructed $H$ against $z$ (see Fig. (\ref{recc})) and the plot shows the universe is expanding. We also get reconstructed bulk viscous pressure $\Pi$ (see Eq.(\ref{ppp})), $\xi$ (see Eq.(\ref{xxx})), density $\rho_{\Lambda}$ (see Eq.(\ref{dens})), $w_{\Lambda}$ (see Eq. (\ref{www})) and squared speed of sound ${v_s}^2$ (see Eq.(\ref{vvv})). We plot ${v_s}^2$ against $z$ (see Fig.\ref{sss})and the plot is feasible. As ${v_s}^2>0$ which indicates the stability of the model. We plotted $w_{\Lambda}$ against $z$ (see Fig.(\ref{bb})) and we observe that $w_{\Lambda}\leq-1$ indicating that the model is phantom but not crossing the phantom boundary.

The reconstructed EoS parameter for different combinations of $\xi_0$, $\xi_1$ and $\xi_2$ have been computed for the current universe ($z=0$) and presented in Table\ref{tab1} and Table\ref{tab2} for non-interacting and interacting scenario respectively. Comparing the values with the values of EoS by the observational scheme Plank + WP+ BAO i.e. $\text{-1.13}^{+0.24}_{-0.25}$, we observe that the reconstructed EoS parameter $p$ is consistent with the observation in both the cases \cite{ww01}.

Next, we choose presence of dark matter in interacting scenario in presence of bulk-viscosity.So, we take $\rho_m = \rho_{m0} a^{-3(1-\delta)}$, we get reconstructed $H$ (see Eq.(\ref{adf})). We plotted $H$ against $z$ (see Fig. (\ref{rq}))and the plot shows that universe is expanding. We then derived reconstructed density $\rho_{\Lambda}$ (see Eq.(\ref{re})) bulk viscous pressure $\Pi$ (see Eq.(\ref{rw})), $\xi$ (see Eq.(\ref{rc})), $w_{\Lambda}$ (see Eq.(\ref{rx})) and squared speed of sound ${v_s}^2$ (see Eq.(\ref{xk})). We plotted ${v_s}^2$ against $z$ (see Fig.\ref{km}) and this plot is  feasible. As ${v_s}^2>0$ which indicates the stability of the model. We plotted $w_{\Lambda}$ against $z$ (see Fig.(\ref{j})) and we observe that $w_{\Lambda}\leq-1$ indicating that the model is phantom but not crossing the phantom boundary.

Next we studied background evolution in viscous interacting DE as scalar field and we calculated $\epsilon_V$ (see Eq.(\ref{epv})), $\eta_V$ (see Eq.(\ref{etv})), $\epsilon_H$ (see Eq.(\ref{abc})) and plotted $\epsilon_H$ against $z$ (see Fig.\ref{ep})  and we can see from the plot  that the universe is expanding and calculated $\eta_H$ (see Eq.(\ref{ett})) and plotted $\eta_H$ against $z$ (see Fig.\ref{ettp}). From the Fig.\ref{ep} and Fig.\ref{ettp} we see that the Hubble slow roll parameters. Therefore, this model  $\epsilon_H$ and $\eta_H$ are increasing respectively. Therefore, this model has the scope of exit from inflation. We calculated tensor to scalar ratio '$r$'(see Eq.(\ref{ts})) and plotted $r$ against $z$ (see Fig.\ref{r}). It is observed that $r<0.168$ which is consistent with the data by Plank Satellite. Hence it can explain the primordial fluctuation in the early universe. We calculated $2 V - {\dot{\phi}}^2$ (see Eq.(\ref{pott})) and plotted $2V-\dot{\phi}^2$ against $z$ (see Fig.\ref{pot}). We have observed that $2V-\dot{\phi}^2>0$, so it can be interpreted that quasi-exponential expansion is available for the interacting viscous HDE with higher order derivative. Moreover, increase in the strength of interaction lowers the rate of expansion in presence of bulk-viscosity.We calculated $w_{\phi}$ (see Eq.(\ref{wph})) and plotted $w_{\phi}$ (see Fig.\ref{wphh}).We observed that $w_{\phi}>-1$, hence it is quintessence. We calculated $\Gamma$ (see Eq.(\ref{game})) and plotted $\mid\Gamma\mid$ vs $z$ (see Fig.\ref{gamee}). We observed that $\mid\Gamma\mid>1$, this implies that it is warm inflation. Plotting $\mid\frac{\Gamma}{3H}\mid$ against $n_s$ (see Fig.\ref{games}), we observed that $\mid\frac{\Gamma}{3H}\mid>1$, hence it is strong dissipative regime.

\textbf{Acknowledgement}: The authors express sincere thanks to the anonymous reviewers for the constructive suggestions. Surajit Chattopadhyay acknowledges financial support under the CSIR (Govt of India) Grant Number: 03(1420)/18/EMR-II.

\end{document}